\documentclass[journal]{IEEEtran}
\usepackage{setspace}
\usepackage[T1]{fontenc}
\usepackage{cite}
\usepackage{caption}
\usepackage{algcompatible}
\usepackage{mathtools}
\usepackage{float}
\usepackage{stackengine}

\usepackage{amsmath,amssymb,amsfonts}
\usepackage{amsmath,epsfig,cite,amsfonts,amssymb,psfrag,subfigure}
\usepackage{graphicx}
\usepackage{blindtext}
\usepackage{textcomp}
\usepackage[dvipsnames]{xcolor}
\usepackage{algorithm}
\usepackage{algorithmicx}
\usepackage[noend]{algpseudocode}
\usepackage{amsthm}
\def\BibTeX{{\rm B\kern-.05em{\sc i\kern-.025em b}\kern-.08em
    T\kern-.1667em\lower.7ex\hbox{E}\kern-.125emX}}
\usepackage{balance}
\allowdisplaybreaks

\usepackage[margin=0.7in]{geometry}
\usepackage[export]{adjustbox}
\usepackage{booktabs,siunitx}
\usepackage{tabularray}

\begin{document}

\title{ Towards Secure Intelligent O-RAN Architecture: Vulnerabilities,
Threats and Promising Technical Solutions using LLMs  \vspace{-.1cm}
}
\author{\small Mojdeh Karbalaee Motalleb$^{\dagger}$, Chafika Benza\"{i}d$^*$, Tarik Taleb$^*$$^+$, Marcos Katz$^*$, Vahid Shah-Mansouri$^\dagger$, JaeSeung Song$^+$ \\
\IEEEauthorblockA{$^\dagger$School of ECE, University of Tehran, Tehran, Iran \\  
$^*$School of CWC, University of Oulu, Oulu, Finland\\
$^+$ Department of Computer and Information Security, Sejong University} 
\\ Email: \{mojdeh.karbalaee, vmansouri\}@ut.ac.ir ,\{chafika.benzaid, tarik.taleb, marcos.katz\}@oulu.fi,\{jssong\}@sejong.ac.kr }
\maketitle
\pagenumbering{gobble}
\begin{abstract}
The evolution of wireless communication systems will be fundamentally impacted by an open radio access network (O-RAN), a new concept defining an intelligent architecture with enhanced flexibility, openness, and the ability to slice services more efficiently. For all its promises, and like any technological advancement, O-RAN is not without risks that need to be carefully assessed and properly addressed to accelerate its wide adoption in future mobile networks. In this paper, we present an in-depth security analysis of the O-RAN architecture, discussing the potential threats that may arise in the different O-RAN architecture layers and their impact on the Confidentiality, Integrity, and Availability (CIA) triad. We also promote the potential of zero trust, Moving Target Defense (MTD), blockchain, and large language models~(LLM) technologies in fortifying O-RAN's security posture. Furthermore, we numerically demonstrate the effectiveness of MTD in empowering robust deep reinforcement learning methods for dynamic network slice admission control in the O-RAN architecture. Moreover, we examine the effect of explainable AI (XAI) based on LLMs in securing the system.
\end{abstract}

\section{Introduction}
\noindent
Wireless systems are becoming more capable but more complex in the next generation of cellular networks. Unlike previous generations, the next generation will be flexible, agile, modular, supporting heterogeneity in services, multiple technologies, and rapid deployment~\cite{polese2022understanding}. 
Radio access networks~(RAN) performance is expected to be significantly improved with O-RAN, which combines and evolves the cloud RAN~(C-RAN) and virtual RAN~(vRAN) to enable an open and flexible RAN. 
In the O-RAN architecture, the components of RANs are virtualized and decoupled, using compatible open interfaces developed for their interconnection.
Moreover, the O-RAN's architecture utilizes artificial intelligence and machine learning~(AI/ML) techniques to develop intelligent RAN layers, allowing to empower intelligent, data-driven closed-loop control for the RAN~\cite{oranSlice}. These features bring many benefits to the system, including reduced capital expenditures~(CAPEX) and operating expenses~(OPEX), increased agility and flexibility, and enhanced visibility and security. 

For all its promises, and like any technological advancement, O-RAN is not without risks that need to be assessed and properly addressed to accelerate its wide adoption in future mobile networks. Indeed, recent studies have shown that the O-RAN architecture is opening the door to a new breed of security challenges brought by the new components and open interfaces defined, the use of open-source software, the disaggregation between hardware and software, and the reliance on cloud-native and AI technologies, among others~\cite{mimran2022evaluating}. Thus, a review of the security aspects needs to be carried out, considering the potential risks and vulnerabilities, as well as the concrete solutions to apply. Such an investigation is essential to strengthen the security posture of O-RAN at its early stage of development.

This paper explores security threats across layers of the intelligent O-RAN architecture and proposes key technologies to mitigate them, highlighting the need for proactive measures in securing next-generation networks \cite{oranSlice}. Unlike prior studies, our research focuses on diverse vulnerabilities in O-RAN, offering an innovative solution for securing near-Real-Time RAN Intelligent Controller (near-RT RIC) and non-Real-Time RAN Intelligent Controller (non-RT RIC) that integrate AI/ML methods for system automation, safeguarding AI/ML models against various potential threats\cite{polese2022understanding, motalleb2023moving}.
Moreover, the near-RT RIC and non-RT RIC includes third party applications which can use AI/ML techniques for the resource allocation.

In addition to traditional security mechanisms, we also propose the novel use of Large Language Models~(LLMs) to enhance the system's security, particularly.
The LLM system can analyze data and articulate the situation in human-readable language to assist in detecting vulnerabilities within the system. The LLM model can use explainable AI (XAI) to analyze the data pattern and realize if there are any significant changes during the time and warn of the vulnerabilities.

Research contributions of this paper are listed as follows: 
\begin{itemize}
    \item An in-depth analysis of vulnerabilities and threats in the O-RAN architecture arising from the introduction of new technologies and common 5G RAN security issues.
    \item The proposal of three countermeasure approaches utilizing the zero trust concept, blockchain technology, and the LLM \& MTD paradigm.
    \item Case studies and proof-of-concept demonstrations of MTD-based robust ML in O-RAN and LLM-based robust AI/ML in O-RAN, illustrating the effectiveness of MTD in enhancing the robustness of deep reinforcement learning models.
\end{itemize}
The remainder of this paper is as follows: Section II provides an overview of the O-RAN architecture, focusing on its key components: RAN, cloud, and management layers, along with ML and network slicing. Section III examines vulnerabilities and threats in the O-RAN architecture, analyzing their impact on confidentiality, integrity, and availability (CIA). Section IV explores emerging technologies such as zero-trust (ZT), blockchain, moving target defense (MTD), and LLMs to enhance O-RAN security. In Section V, we propose a novel MTD-based solution demonstrating its effectiveness in securing deep reinforcement learning (DRL) against adversarial attacks in the Near-RT RIC. Additionally, we discuss the application of LLM-based explainable AI (XAI) for detecting AI/ML attacks in O-RAN. Finally, conclusions are drawn in Section \ref{sec:conclusion}.
\section{O-RAN Background}\label{ORANB}

\noindent
The O-RAN Alliance\footnote{https://www.o-ran.org} has developed a novel RAN architecture to facilitate an open, intelligent, virtualized, and interoperable RAN, essential for cost-effective, next-generation wireless networks. This architecture integrates the advantages of C-RAN and vRAN, leveraging cloudification, centralization, and hardware-software decoupling to address vendor lock-in and proprietary issues via standard interfaces. O-RAN developed a multi-vendor ecosystem and embedded AI/ML for improved network intelligence.

\begin{figure}
\vspace{-1.5em}
\centering
\subfigure[]{
\includegraphics[scale=0.28, trim={0cm 0 0 0}]{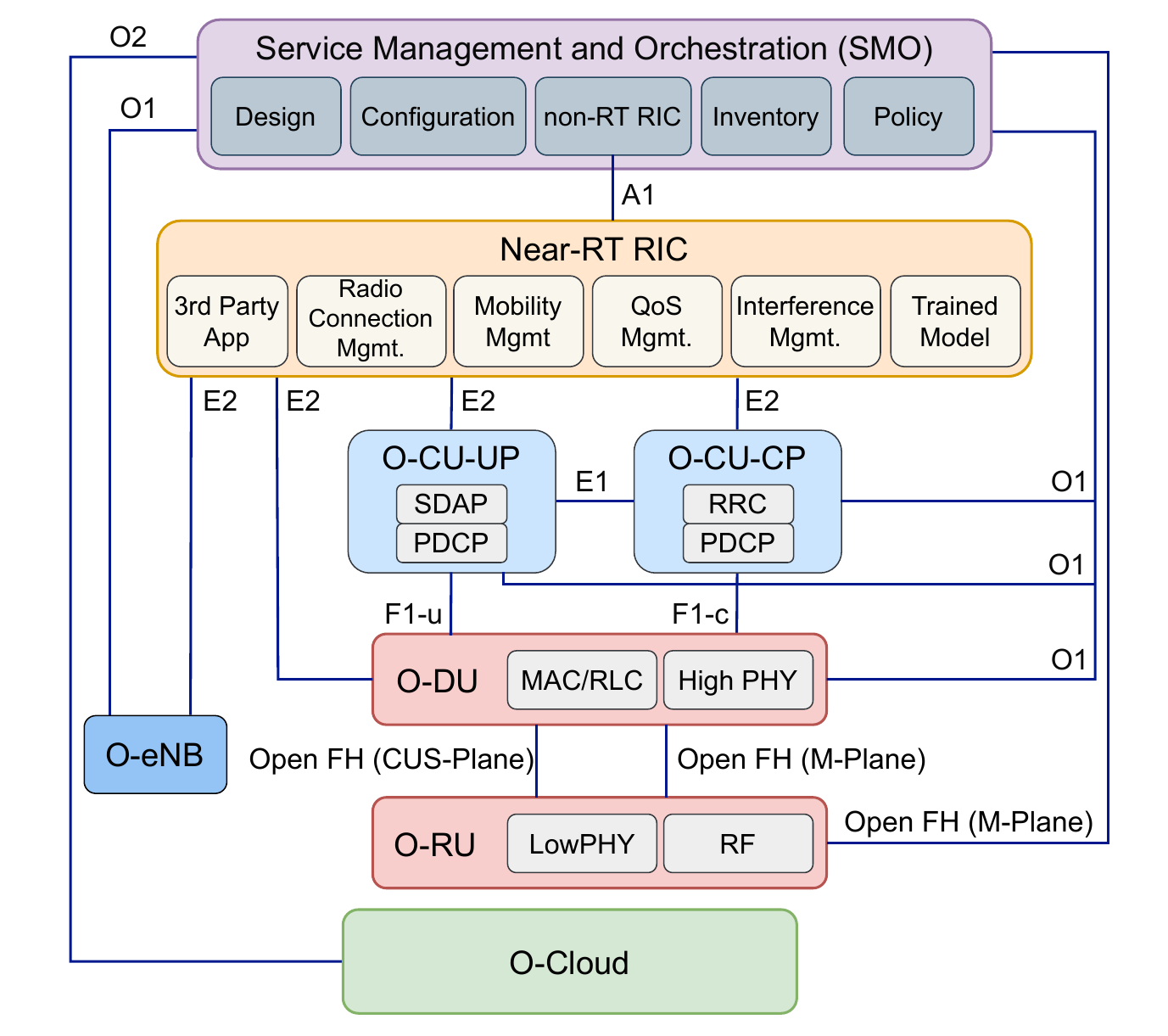}\label{fig:c11}
}\hspace{-0.25cm}
\subfigure[]{
\includegraphics[scale=0.28, trim={0 0 0 0}]{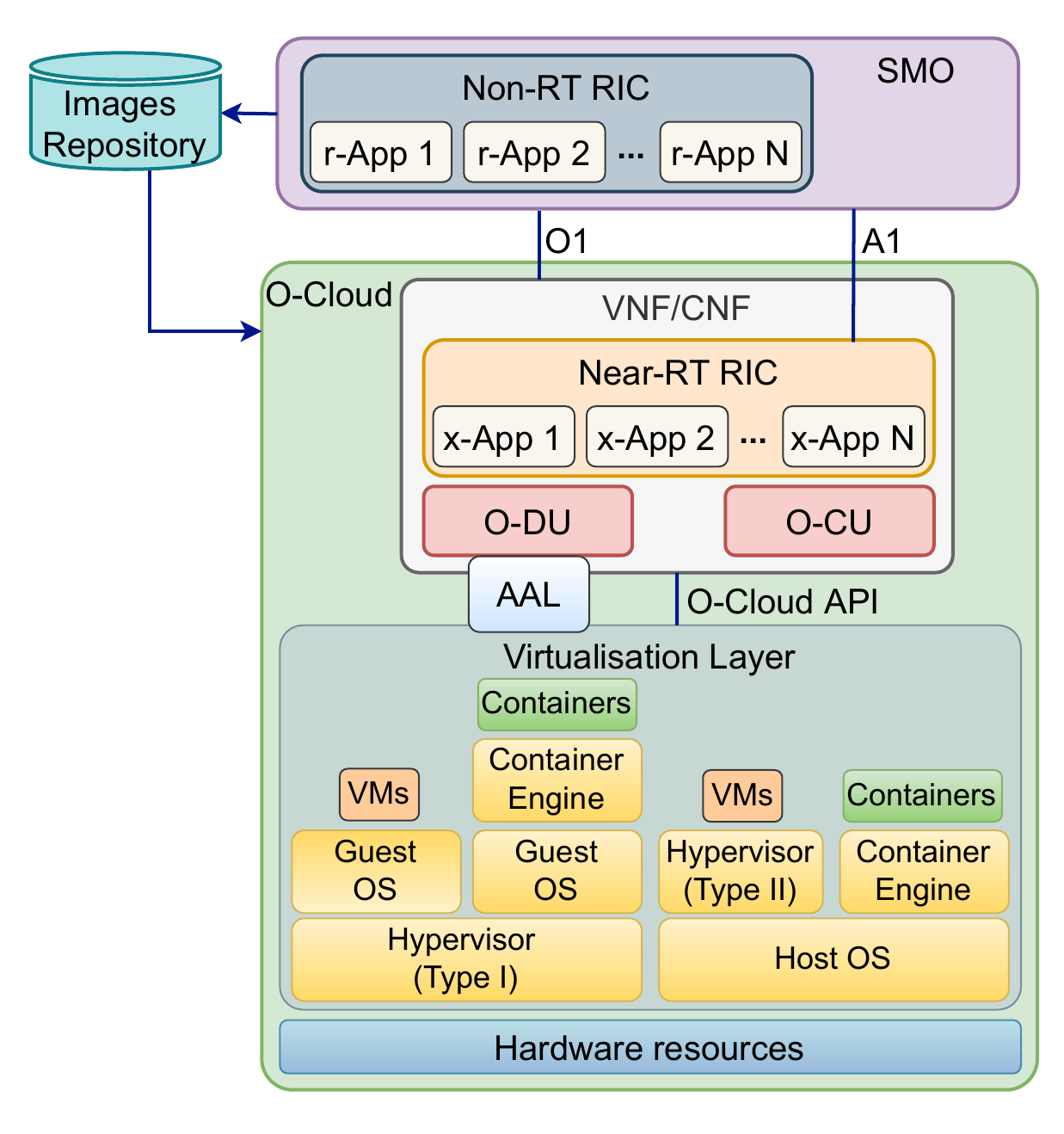}\label{fig:c12}
}
\caption{(a) The O-RAN high-level architecture with components and interfaces, (b) The O-Cloud architecture, which is a set of computing resources and virtualization infrastructure.}
\label{fig:c14}
\end{figure}
\textcolor{black}{
The O-RAN architecture includes three components in the baseband side: the Radio Unit (O-RU), Distributed Unit (O-DU), and Central Unit (O-CU).
The O-RU contains the radio frequency (RF) and low physical (PHY) layers, while O-DU provides the functionalities of the high PHY, Medium Access Control (MAC), and Radio Link Control (RLC) layers.
The open fronthaul (Open-FH) is the interface between the O-RU and the O-DU. The Open-FH interface includes a control user synchronization plane (CUS-plane) and a management plane (M-plane).
The O-CU is divided into two logical nodes the user plane (O-CU-UP) and the control plane (O-CU-CP). The O-CU-UP encompasses the service data adaptation protocol (SDAP),
and the user plane part of the packet data convergence protocol (PDCP).
The O-CU-CP hosts the radio resource control (RRC) layer,
and the control plane of the PDCP protocol.
Fig.\ref{fig:c11} illustrates O-RAN's architecture.}

\textcolor{black}{The O-RAN architecture also includes a management part which comprises Service management and Orchestration (SMO), Near Real-Time RAN Intelligent Controllers (RICs), and O-Clouds blocks. SMO includes functions such as Non-Real-Time RIC. Generally, the near-RT and non-RT RIC are responsible for AI/ML methods and making the system more intelligent. The AI/ML technologies plays a crucial role in the resource
allocation within RAN systems.
In the O-RAN system, near-RT RICs are functions that provide near real-time control and optimization of network resources through the E2 interface. This includes xApplications (xApps), which are third-party applications that run by leveraging the modules and capabilities of a system for functionalities such as resource allocation. }

\textcolor{black}{The O-Cloud platform, known as a cloud computing platform, hosts O-RAN architecture components depicted in Fig.~\ref{fig:c12} \cite{ORANSecOcloud}.
The RAN network functions can be deployed as virtualized network functions (VNFs) on virtual machines (VMs) or as cloud-native network functions (CNFs) in containers. The O-Cloud platform supports these options with its virtualization layer, which includes operating systems, hypervisors, and container engines. Additionally, the O-RAN ecosystem supports and interfaces with bare-metal, hardware-based RAN functions. The SMO system connects to the O-Cloud via the O2 interface, enabling efficient resource and workload management \cite{ORANArch}.}

In the following, we provide a concise overview of the key techniques and features employed within the O-RAN system, enhancing its flexibility and performance.
\subsection{Network Slicing in O-RAN}
Network slicing, essential for 5G revenue, dynamically creates customized virtual networks on shared infrastructure, integrating network functions and resources across RAN, transport, and core networks to meet specific service needs.
RAN slicing involves the isolation of Physical Resource Blocks (PRBs) and specific Virtual Network Functions (VNFs) such as MAC, RLC in the O-DU, and PDCP, SDAP in the O-CU for various services as illustrated in Figure 1 of \cite{oranSlice}.
In addition, core slicing virtualizes and isolates nodes like UPF and AMF, catering to the specific needs of each service.
Finally, transport slicing creates dedicated pathways across the shared underlay network, ensuring guaranteed performance for these diverse service connections. By working together, RAN, core, and transport slicing unlock the full potential of 5G networks.
O-RAN's virtualization and intelligence are key to advancing RAN slicing, essential for end-to-end network services \cite{oranSlice}. 

\subsection{Radio Intelligent Controller (RIC)}
The Near-RT and Non-RT RICs are essential for O-RAN system management, serving as an open hosting platform and optimizing RAN functions.
The RIC consists of Near-RT RIC and Non-RT RIC, facilitating intelligent RAN optimization on near-real-time ($10-1000$ msec) and non-real-time (greater than $1$s) scales, respectively. The Near-RT RIC uses xApps for real-time RAN control via E2 interfaces with O-RAN components, while the Non-RT RIC employs rApps for broader RAN optimization and is linked to the Near-RT RIC through the A1 interface for policy and AI/ML model management. 
The near-RT RIC and non-RT RIC are vital components responsible for the AI/ML workflow in the O-RAN architecture\cite{ORANArch, ORANML,polese2022understanding}.

\subsection{ML aspect in O-RAN}
The O-RAN architecture incorporates AI/ML to add intelligence across its RAN layers, a move seen as pivotal for highly autonomous RAN functions that improve service quality and lower OPEX. AI/ML is expected to be instrumental in a range of RAN use cases, from resource allocation to anomaly detection and cybersecurity. Subsequently, we will outline potential ML techniques applicable to O-RAN and detail the general ML lifecycle.

\subsubsection{ML techniques}
In the O-RAN system, various ML techniques are utilized: (1) supervised learning for model training with labeled data and subsequent prediction on new data; (2) unsupervised learning to find patterns in unlabeled data; (3) reinforcement learning (RL) and deep RL (DRL) for learning optimal actions through interaction with the environment; and (4) federated learning (FL) for privacy-preserving collaborative model training across distributed entities without data exchange, using a central server to aggregate local model updates.
In addition, LLMs can also be incorporated to enhance communication performance and the decision-making processes by analyzing and generating human-like text, providing valuable insights within the O-RAN architecture. Moreover, integrating LLMs with existing ML methods can significantly improve the system's overall intelligence and efficiency.

In O-RAN architecture, Non-RT RIC and Near-RT RIC are responsible for AI/ML techniques, where they can play the role of ML training host and/or ML model host/actor \cite{ORANML}. The ML training host VNF trains models within the Non-RT RIC, while the ML model host/actor VNF, for inference, may reside in either Non-RT or Near-RT RIC. 
In RL, Near-RT RIC conducts online training and inference, while Non-RT RIC is for offline training and Near-RT RIC for inference. FL uses Non-RT RIC as the central server and Near-RT RIC for distributed training.
\subsubsection{ML Life Cycle Procedure}
\begin{figure}
  \centering
  \resizebox{0.4\textwidth}{!}{
    \includegraphics{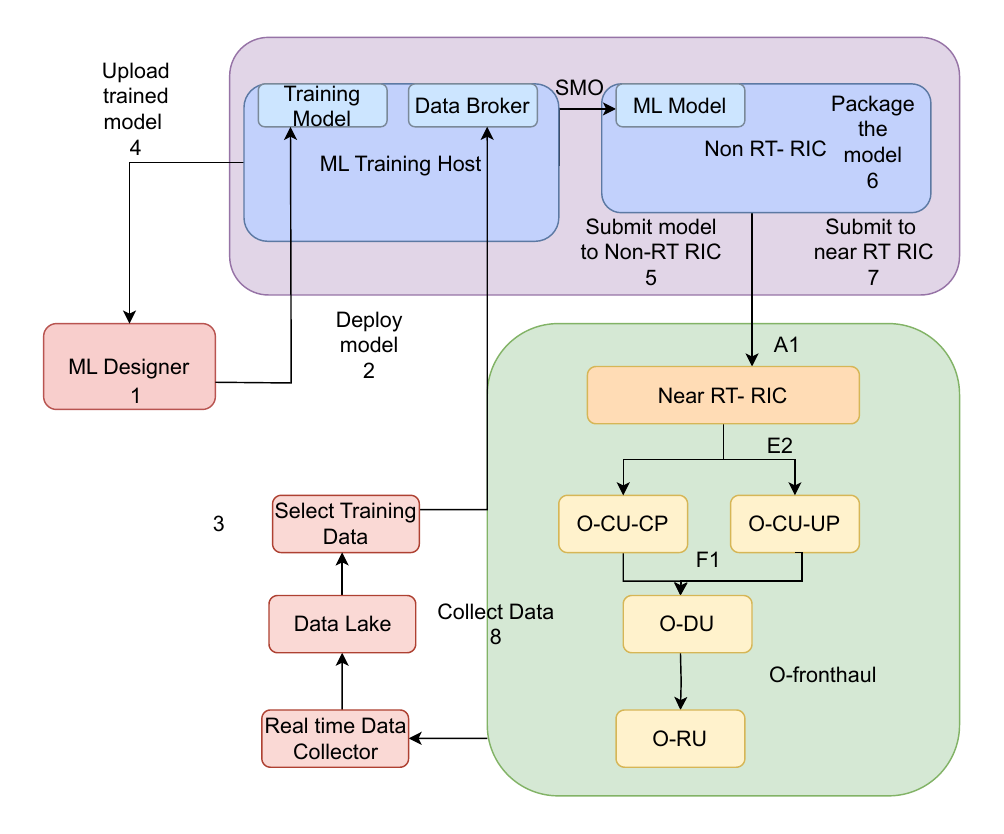}
  }
  \caption{ML Model Life Cycle in the O-RAN Architecture.}
  \label{fig:ml}
\end{figure}

\textcolor{black}{
Despite the variety of ML techniques supported and the deployment scenarios considered for placing the ML training hosts and ML model hosts/actors, a general ML lifecycle in the O-RAN architecture can be described as follows (See Fig. \ref{fig:ml}) \cite{polese2022understanding,ORANML}:Firstly, the ML Designer, deoployed the model (stage 1 and 2). The data is selected for training (stage 3) and fed into the ML model during the training and inference stages. The data are typically collected over E2, O1, and A1, from O-CU, O-DU, and RICs (stage 8). The collected data are prepared in the RICs to fit the ML models by performing data pre-processing operations, including dataset balancing, normalization, and removing noise, among others. The ML model goes first through the training process, where the ML designer or SMO/Non-RT RIC will select and implement the ML algorithm to train in the ML training host. The trained model is then uploaded (stage 4) and validated to ensure its reliability and accuracy. Once the model is validated, it is stored and published in the SMO/Non-RT RIC catalog (stage 5). After a model has been validated (stage 6), it can be deployed and executed (stage 7).}
\section{Vulnerabilities and Threats in O-RAN Architecture}\label{vulner}
\noindent
O-RAN architecture's openness and disaggregation facilitate compliance with security standards and enable improved security agility, adaptability, and resiliency for future mobile networks. In addition to those benefits, O-RAN architecture introduces the potential for an increased attack surface \cite{ORANSec}.
The O-RAN Alliance's Security Work Group 11 focuses on securing O-RAN, but their measures are insufficient, particularly against malicious AI/ML methods. Therefore, additional security perspectives are necessary. This section discusses key vulnerabilities and threats to O-RAN, including the new O-RAN technologies security issues.

\subsection{O-RAN System Vulnerabilities}
As previously discussed, the O-RAN system comprises three different sides (radio, management, cloud), each with its own vulnerabilities tied to their respective roles and functions. This section delves into the vulnerabilities inherent to the different sides of the O-RAN architecture.

\subsubsection{O-RU/O-DU and Open-FH Vulnerabilities}
In radio communication, the O-RAN architecture and other RAN generations have inherent vulnerabilities. This section outlines these vulnerabilities, particularly focusing on O-RAN. One key threat is the false base station (FBS) attack, where an attacker poses as a legitimate base station to execute a Man-in-the-Middle (MiTM) attack. Three FBS attack scenarios on an O-RU include hijacking fronthaul, recruiting a standalone O-RU, and gaining unauthorized physical access. These attacks can compromise both O-RAN and other RAN systems~\cite{ORANSec}.

There are several risks associated with FBSs in the network, including stealing subscriber information, altering and redirecting transmitted data, and compromising subscriber privacy. The FBS attacks may help in penetrating O-DU and beyond in the CN and launching DoS attacks to cause loss of service or degradation of its performance.

Given that the O-DU and O-RU can be from different vendors, they may have varying security levels. The O-DU's role in managing traffic between the management system and the O-RU increases the risk of unauthorized access to other systems, such as RICs, via the Open-FH interface. An unprotected Open-FH interface can also enable Man-in-the-Middle (MiTM) attacks, allowing data tampering, disclosure, and DoS attacks. For instance, an unauthorized device on the Open-FH Ethernet L1 interface could launch a flooding attack, causing unavailability or performance degradation of legitimate network elements.
\subsubsection{Near-RT RIC Vulnerabilities}
Through standardized interfaces and hardware support, the Near-RT RIC provides a safe and reliable platform for hosting xApps. The xApps are independent of the Near-RT RIC and may be supplied by a third-party vendor. The Near-RT RIC and xApps can be sources of different security threats \cite{ORANSec}. 

A malicious or compromised xApp has the potential to negatively impact the service delivery for a subscriber, a group of subscribers, or a specific geographic area by manipulating data collected from E2 nodes (i.e., O-DU, O-CU-CP and O-CU-UP) and A1 interface. It introduces also the risk of obtaining unauthorized access to E2 nodes and Near-RT RIC, exploiting the RAN functions and engendering harmful effects to the overall system. Leakage of sensitive data (e.g., UE identification and location) is another menace that could stem from malicious/compromised xApps. The disclosure of sensitive information will not only pose privacy violation issues but may also lead to the launch of other attacks, such as impersonation and UE tracking attacks. 
The xApps cannot operate independently from the components of the Near-RT RIC. They need to interact with these components to access their functionalities. For instance, they communicate with the App Manager during registration and the Sub Manager to subscribe to data from E2 nodes. Due to this communication, a malicious xApp can affect other components of Near-RT RIC too. 
\textcolor{black}{This could happen by exploiting shared resources, manipulating control messages, disrupting event processing, compromising security credentials, introducing hidden logic bombs, or exfiltrating sensitive data through communication channels within the framework.}
Additionally, resources such as CPU and RAM limits can be specified in the xApp descriptors to prevent resource exhaustion, which is enforced by Kubernetes. Hence, a malicious xApp can use more resources than it needs.

The indefinite functional split between Near-RT RIC and E2 nodes, which depends on the available xApps and the capabilities of E2 nodes, may result in conflicts between decisions taken by the Near-RT RIC and the E2 nodes. Moreover, developing multiple xApps with overlapping objectives within the same RAN may lead to conflicting actions between xApps. Those conflicts can degrade the system's performance or may cause a Denial-of-Service (DoS) attack intentionally or unintentionally in the O-RAN architecture.

The lack of proper isolation between an xApp and the other Near-RT RIC components may be a source of serious security breaches. In fact, with the recent trend to evolve VNFs into CNFs, complete isolation between co-hosted CNFs is hard to realize due to the lack of strong hardware isolation in the emerging cloud-native platforms (e.g., Kubernetes). 
Thus, an xApp with compromised isolation can be exploited to escalate the privilege granted to it, carry out shared resource exhaustion attacks, steal secrets and sensitive information from memory, and conduct DoS attacks against co-hosted xApps and the Near-RT RIC platform.

\subsubsection{SMO Vulnerabilities}
SMO security is critical because a vulnerability can allow attacks on O-RAN components and lateral movement within the network. Weak authentication and authorization can let attackers access and alter SMO data, control O-RAN components, and steal sensitive information. For example, unauthorized access to Non-RT RIC via SMO can lead to UE tracking or issuing false policies to Near-RT RIC. Additionally, SMO and Non-RT RIC are susceptible to DoS attacks, which can impair network monitoring and control functions. The security concerns for rApps in Non-RT RIC are similar to those for xApps~\cite{ORANSec}.

\subsection{O-Cloud Vulnerabilities}
The O-Cloud platform in O-RAN architecture faces common cloud security risks, including software flaws, valid account access, and lack of interface authentication. Malicious actors can exploit VMs and containers running O-RAN components, leading to privilege escalation, malware contamination, unauthorized deployment of VMs/containers, root server access, and system destruction. They can also access and manipulate sensitive data. Deploying vulnerable VMs/containers risks DoS attacks on shared resources, which can be economically damaging if turned into an EDoS attack. Supply chain attacks can inject malicious code or extract private keys from VM/container images. Additionally, an unprotected O2 interface between O-Cloud and SMO is vulnerable to MiTM attacks, allowing tampering and disclosure of services and requests.

\subsection{Open Source Code Vulnerabilities}
Open-source software is crucial for building the software-based O-RAN architecture, used in both cloud infrastructure and O-RAN components. It accelerates development, promotes vendor independence, and reduces costs. However, it also poses security challenges. The open source code allows attackers to find and exploit vulnerabilities. Without an accurate, up-to-date inventory of open-source codes and dependencies, managing and mitigating high-risk vulnerabilities becomes difficult due to the volume, variety, and lack of standard naming conventions.
 
\subsection{ML System Vulnerabilities}
Integrating ML techniques into O-RAN enhances autonomous RAN functions but also introduces significant security challenges. ML models are vulnerable to adversarial attacks that manipulate decisions, compromise model integrity, or reveal private information. Attacks include altering training datasets, injecting fake data during online learning, or crafting inputs to deceive models during operation. Collaborative learning methods like FL face model poisoning attacks, where malicious agents tamper with local model parameters to compromise the global model. FL is also susceptible to inference attacks, allowing attackers to deduce private training data using local model parameters~\cite{mimran2022evaluating, aisecme}.

Based on accessibility, attacks on ML models can be categorized into white-box, black-box, and gray-box attacks~\cite{aisecme}. Indeed, the adversarial attack is considered as a white box, gray box, or black box when the attacker can have full, partial, or no access to the training data and the targeted model's parameters and architecture, respectively. The white-box attack is deemed less realistic due to the assumption of an attacker with full knowledge, which is hard to achieve in real-world scenarios.
 
 \subsection{Threats against 5G Radio Networks}
Common threats to traditional RAN architectures are also applicable to O-RAN architecture. This includes (i) jamming attacks, which consist of blocking radio signals; for example by introducing intentional interference in the communication channels; (ii) sniffing attacks, which focus on observing and collecting data packets with the purpose of extracting sensitive information (e.g., UE location and cell configuration) as well as using the extract information to craft new attacks; and (iii) spoofing attacks, which refer to creating a fake signal that is hard to distinguish from the actual signal, allowing an attacker to impersonate a base station, cause a DoS, or bypass physical-layer signal authentication~\cite{aisecme}, among others.

\setlength{\tabcolsep}{0.17em}
\begin{table*}
\vspace*{-0em}
 \caption {Impact of threats and vulnerabilities in O-RAN system on Confidentiality (\textbf{C}), Integrity (\textbf{I}) and Availability (\textbf{A}); and the Potential Mitigation of Vulnerabilities through Zero Trust (\textbf{ZT}), Blockchain (\textbf{BC}), Moving Target Defense (\textbf{MTD}),  Large Language Model (\textbf{LLM}).} \label{table:1}
 \begin{center}
 \scalebox{.8}{
 
\centering
\begin{tblr}
{|l | c c c| c c c c|}
  \hline
\textbf{Threats and Vulnerabilities} & \textbf{C} & \textbf{I} & \textbf{A} & \textbf{ZT} & \textbf{BC} & \textbf{MTD} & \textbf{LLM}\\ [0.5ex]
   \hline
  Conflicts among xApps or rApps  & x & x & \checkmark  & x & x & \checkmark & \checkmark\\[.5ex]
  \hline
  Accessing a misconfigured x/rApps  & \checkmark & x & x & x & \checkmark & x & \checkmark \\[.5ex]
  \hline
  Altering Data through malicious x/rApps attacks & \checkmark & \checkmark & x  & \checkmark & \checkmark & x & \checkmark\\[.5ex]
  \hline 
  Conflicts between Near-RT RIC and O-gNB/eNB  & x & x & \checkmark   & x & x & \checkmark & \checkmark\\[.5ex]
  \hline
  FBS attacks on O-RU & \checkmark & \checkmark & x  & \checkmark & \checkmark & x & x\\[.5ex]
  \hline
  Eavesdropping on air interfaces & \checkmark & x & x  & x & x & x & x\\[.5ex]
  \hline
  Accessing the O-RU/DU/CU and degrading the O-RAN's performance  & x & x & \checkmark & x & x & \checkmark & x\\[.5ex]
  \hline
  MITM attack from the Open-FH over M-plane or CUS-plane  & \checkmark & \checkmark & x & \checkmark & \checkmark & x & x\\[.5ex]
  \hline
  Misconfiguration, lack of isolation and security in the O-Cloud  & \checkmark & \checkmark & \checkmark & \checkmark & \checkmark &  \checkmark  & \checkmark\\[.5ex]
  \hline
  Open-source code vulnerabilities & \checkmark & \checkmark & \checkmark & \checkmark & \checkmark &  \checkmark  & \checkmark\\[.5ex]
  \hline
  Adversarial attacks against ML  & \checkmark & \checkmark & \checkmark & \checkmark & \checkmark &  \checkmark  & \checkmark\\[.5ex]
  \hline
  Jamming attacks  & x & x & \checkmark  & x & x & \checkmark& \checkmark\\[.5ex]
  \hline
  Spoofing attacks & \checkmark & \checkmark & \checkmark  & \checkmark & \checkmark & \checkmark & \checkmark\\[.5ex] \hline 
  Physical threats  & \checkmark & \checkmark & \checkmark  & x & x & x & x\\[.5ex]
  \hline
\end{tblr}
}
 \end{center}
 \end{table*}

\subsection{Physical Threats}
Physical threats, though not unique to O-RAN, are crucial to understanding its vulnerabilities. The physical infrastructure, including cell sites and data centers, faces risks from unauthorized access, power outages, natural disasters, and hardware failures. Intruders can sabotage hardware or alter settings to provoke DoS, inject malware, or access other network components. Natural disasters like snow, floods, earthquakes, and lightning can damage physical components. Lack of proper procedures for hardware failures and power outages increases the risk of unavailability. Physical security is more challenging in O-RAN due to the higher number of cell sites, data centers, and vendors.

Table~\ref{table:1} summarizes the main security threats discussed above, highlighting their impact on the CIA triad. Note that the threats marked with the (\checkmark) sign affect a CIA principle, while those marked with (x) do not. Moreover, (\checkmark) and (x) indicate whether the potential mitigation of vulnerabilities through Zero Trust (ZT), Blockchain (BC), Moving Target Defense (MTD), and LLM investigated in Section \ref{security} is applicable or not, respectively.

\section{Security Solutions in O-RAN}
\label{security}
\noindent
There are different possible solutions for security threats and vulnerabilities~\cite{trustme}. This section discusses several key emerging technologies that can be leveraged to improve the security of the O-RAN architecture.

\subsection{Zero Trust}
Zero trust (ZT) is a valuable security model for enhancing O-RAN security. Based on "never trust, always verify," it assumes breaches can occur anytime from internal or external threats. ZT principles include continuous identification and authentication, enforcing least-privilege access, maintaining risk-based policies, checking communication channels, and continuous security monitoring. Implementing ZT protects the entire O-RAN architecture, from hardware to applications. AI/ML techniques and Security-as-a-Service (SECaaS) enable ZT by allowing instant threat identification and automated security adjustments~\cite{vnf1}.

\subsection{Blockchain}
Blockchain (BC) is a promising solution for securing O-RAN architecture with a zero trust mindset. Its features of decentralization, immutability, transparency, auditability, and smart contract auto-execution support various security controls in O-RAN. These controls include privacy-enhanced identity management, mutual authentication, dynamic access control, integrity and non-repudiation of data and software, and secure resource sharing. For example, in AI security, blockchain can ensure the integrity and provenance of data in a ML pipeline and protect against poisoning attacks on FL models~\cite{trustme,vnf1}.

\subsection{MTD}
MTD has recently emerged as an effective approach to enable proactive security. The core principle of MTD is to constantly and dynamically modify the configuration of the network and services to increase uncertainty and complexity for attackers. In fact, the dynamicity introduced by MTD reduces the attacker's opportunities to gather useful information on vulnerabilities of the target environment, preventing their exploitation. To this end, different MTD techniques can be applied, which are broadly categorized into \textit{shuffling} (e.g., network topology, VMs/containers placement), \textit{diversity} (e.g., in underlying technology used to implement or run a service), and \textit{redundancy} (e.g., by providing multiple replicas of a network component or service). In O-RAN, the MTD approach can be used to prevent intrusions, mitigate DoS attacks, and increase the robustness of ML models to adversarial attacks (Table~\ref{table:1}), among others. For example, the resiliency of ML models can be strengthened by continuously changing the ML algorithm, the features used for its training, or the model's parameters~\cite{aisecme}. Moreover, to determine whether we have resources to allocate to UE, we can use the AI/ML method for the admission control system. This AI/ML system can be protected using MTD by considering different AI/ML training models with different configurations that are chosen randomly by MTD. 

\subsection{Large Language Models}
The deployment of Large Language Models (LLMs) within O-RAN networks can significantly enhance cybersecurity measures by capitalizing on their exceptional data processing and pattern recognition capabilities. In the context of O-RAN, where a diverse array of virtualized network functions operates across open interfaces, LLMs can meticulously monitor and analyze network traffic and system logs. This enables the early detection of anomalous behaviors that could signal a security breach, such as unusual login patterns or unexpected changes in data flow, which are critical in the multi-vendor O-RAN environment.

LLMs can dynamically adjust security policies for each O-RAN network slice by analyzing data to make smart access choices, fine-tune encryption, and improve intrusion detection, resulting in personalized security.
We can fine-tune LLM system for specific tasks based on our requirements for the next generation of RAN system \cite{lin2023pushing}. For instance, we can fine-tune the LLM system to analyse the data and diagnosis to early warnings.

let's consider a specific scenario: in the event of a sudden surge in traffic indicating a potential DDoS attack within a network slice, an LLM equipped with real-time analytics can autonomously adjust traffic rules and resource allocations to mitigate the threat. This proactive approach not only ensures uninterrupted service but also enhances overall security by continuously monitoring for vulnerabilities and updating configurations.

In the realm of O-RAN, where AI/ML-driven solutions are paramount, LLMs can also contribute to the secure orchestration of network elements by generating and updating security configurations and orchestrating responses to threats in collaboration with the SMO framework. This not only streamlines the management of complex O-RAN architectures but also fortifies them against sophisticated cyber threats, ensuring the network's integrity and the trust of its users.

By integrating LLMs into the O-RAN security strategy, network operators can leverage the full potential of AI to maintain a robust, adaptive, and intelligent defense system, keeping pace with the evolving cyber-security landscape while supporting the continuous growth and innovation inherent to O-RAN networks.

Also, LLMs can be used to enhance XAI systems in O-RAN by providing human-like explanations for the decisions and predictions made by various AI/ML components. As a result, XAI reduces the risk of false positives and improves the accuracy of security AI/ML.
When the systems or operators understand the reasoning behind AI decisions, they can fine-tune the system to be more precise, leading to better detection of genuine threats and fewer mistakes. In other words, XAI with the help of LLMs not only makes AI more transparent but also smarter and more reliable when it comes to keeping the network safe \cite{datta2023s}.

\subsection{Effect of Security Solutions on different Vulnerabilities}
This section examines the impact of ZT, BC, MTD, and LLM on the vulnerabilities listed in Table \ref{table:1}.
Conflicts among xApps or rApps, and between Near-RT RIC and O-gNB/eNB, and accessing the O-RU/DU/CU and degrading the O-RAN's performance can be prevented and resolved by implementing the MTD method which constantly changes the configuration and environment of the system. Moreover, MTD could potentially mitigate jamming attacks by dynamically changing frequencies or communication patterns. 

BC can prevent misconfigured x/rApps from being accessed by ensuring configurations are recorded immutably, making misconfigurations easier to detect.
When malicious x/rApps attacks alter data, BC ensures data integrity, while Zero Trust prevents unauthorized access, mitigating risk. In addition, in order to prevent FBS attacks on O-RU and MITM attacks from the Open-FH over M-plane or CUS-plane, BC could provide a secure and transparent method for firmware distribution and communication channel, while Zero Trust could prevent unauthorized access to the system.

Misconfiguration, open-source code vulnerabilities, and adversarial attacks against machine learning can be secured by employing Blockchain for immutable logging and verification, Zero Trust for rigorous access control and continuous authentication, and MTD to dynamically alter the system's attack surface, complicating potential exploitation efforts.
Moreover, LLM can help in detecting many threats shown in Table \ref{table:1} such as adversarial attacks against AI/ML using XAI, open-source code vulnerabilities, jamming and spoofing using various analyses, and pattern recognitions.

\section{Secure O-RAN Case Studies} 
\label{sec:case-studies}
\noindent
In this section, we investigate two case studies: MTD-based Robust ML and LLM-based XAI Robust AI/ML in O-RAN. The first study explores the application of the MTD approach in enhancing deep reinforcement learning methods for dynamic network slice admission control within the O-RAN architecture. The second study focuses on the use of an LLM XAI system for diagnosing and explaining aberrant behavior.

\subsection{MTD-based Robust ML in O-RAN}
This section presents a practical study corroborating the capabilities of the MTD approach in empowering robust DRL methods for dynamic network slice admission control in the O-RAN architecture~\cite{motalleb2023moving}.
While AI/ML is essential in the O-RAN for functions like resource allocation and network slicing, its security is vital for ensuring the reliability of 5G and 6G networks. Therefore, MTD is chosen for the study due to its agility in reconfiguring ML systems within O-RAN, effectively disrupting attack vectors and fortifying against the complex threats of future wireless networks.

\subsubsection{System Scenario}
\label{serviceAd}
\begin{figure}
  \centering
  \resizebox{0.27\textwidth}{!}{
    \includegraphics{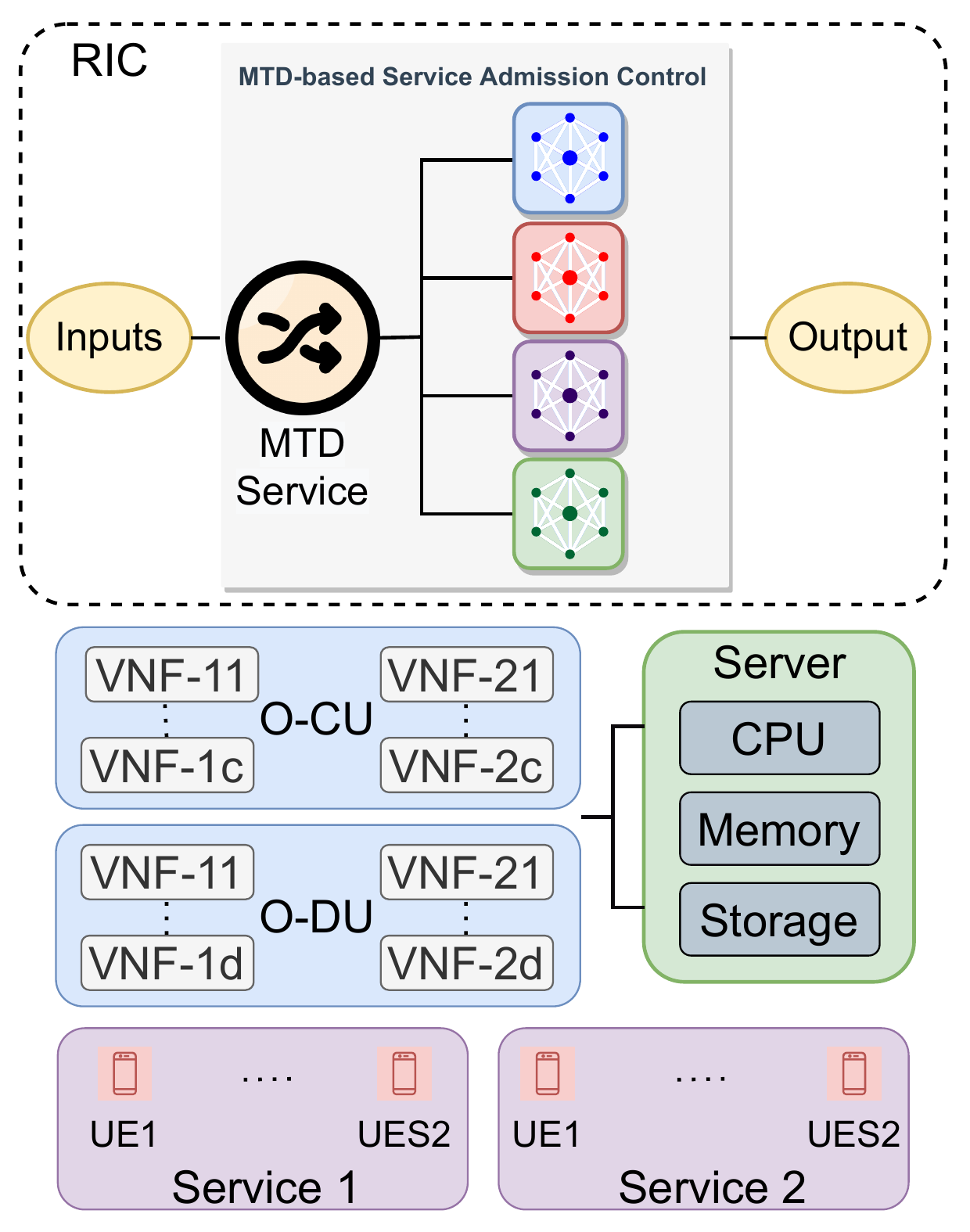}
  }
  \caption{MTD-based dynamic VNF placement scenario based on service request.}
  \label{fig:c15}
\end{figure}

We consider a scenario of service admission control, as shown in Fig. \ref{fig:c15}, in which we have two different services in the O-RAN architecture. In order to provide a service requirement, a specific amount of resources is needed. Each service is assigned to its slices based on the network slicing technique in the O-RAN architecture. Each slice contains VNFs in the O-DU and O-CU layers. 
\textcolor{black}{
In this study, we implement a simulation for the O-RAN architecture by considering the O-DU and O-CU as specific VNFs with memory requirements. For simplicity, we assume that O-DU and O-CU use the same processors. Additionally, in the near-RT RIC, the AI/ML models are trained to solve the resource allocation problem. This model is implemented as an xApp within the system.}
We suppose that the system has enough CPU and storage resources while it has restricted memory resources. We consider a dynamic resource allocation model for VNFs of O-DU and O-CU slices for service admission control problems. 
Our goal is to maximize the total service admission rate.
We suppose that services have the same priority in this system model. 
In this service, we assume the system is dynamic, and in each time slot, we have service requests from the two services that arrive following a Poisson process.
Additionally, we assume that these two services have a service departure rate that has an exponential distribution. 
 
\textcolor{black}{
Suppose we have a tuple that represents the required resources for VNF $m$ in the O-DU or O-CU ($m^\mathfrak{z}$, $\mathfrak{z} \in {c,d}$) within slice $s$, denoted as $\bar{\psi}{s}^{m^\mathfrak{z}} = \{\psi_{\mathsf{C},{s}}^{m^\mathfrak{z}}, \psi_{\mathsf{S},{s}}^{m^\mathfrak{z}}, \psi_{\mathsf{M},{s}}^{m^\mathfrak{z}} \}$. Here, $\psi_{\mathsf{C},{s}}^m$, $\psi_{\mathsf{S},{s}}^m$, $\psi_{\mathsf{B},{s}}^m$, and $\psi_{\mathsf{M},{s}}^m$ indicate the required amounts of CPU, storage, bandwidth, and memory, respectively, for the VNFs of the O-DU (d) or O-CU (c).
}

\textcolor{black}{Assume there are $N$ data centers designated for the VNFs of the O-DU and O-CU. Each data center $n$ possesses a memory resource capacity denoted as ${\chi}_{s}^n$.}

\textcolor{black}{Assume $x_{m^\mathfrak{z}_s,n} \in {0,1}$ is a binary variable indicating whether the VNF $m^\mathfrak{z}_s$ in layer O-DU/O-CU ($\mathfrak{z} \in {c, d}$) within slice $s$ is being hosted by data center $n$.}

\textcolor{black}{In this system model, we aim to maximize the service admission rate ($ \sum_{n=1}^{N}\sum_{m_s=1}^{M_s} x_{m_s,n}$) with the constraint that $x_{m_s,n}$ is a binary variable. Additionally, 
$\sum_{s=1}^{S} \sum_{m_s=1}^{M_s}  x_{m_s,n} \bar{\psi}_{{\mathsf{M}},s}^{\mathfrak{z},tot}  \leq   \chi_{\mathsf{M},{s}}^n \;\; \forall n$, meaning that the total memory used by the VNFs hosted on server n must not exceed the server's total memory.
This problem was modeled and solved in Python using the PPO model which is a DRL method. }

\subsubsection{Proposed Service Admission Algorithm}
\label{proposed}
To solve this service admission control problem, we consider a DRL method that is implemented in the Near-RT RIC. Moreover, we assume the memory is quantized. Therefore, we have discrete action and space. The DRL method adopted is Proximal Policy Optimization (PPO); an actor-critic method. Two models have been developed in the Actor-Critic system, namely: the Actor and the Critic. The Actor decides to take which action, and it updates the policy network for the selected agent. The Critic corresponds to the value function. During updating the Actor, the Critic modifies the network parameters for the value function.
In this system,  the state is the remaining memory we have in each time step, appended to the service arrival rate for two services which are random variables with a Poisson distribution, while the actions are the service admission for the two services. Moreover, the reward is the function of the service admission rate and the remaining memory. A reward is a negative number if the remaining memory is less than zero.

\subsubsection{Attack Model}
\label{attack}
This section describes a malicious adversarial attack on the proposed PPO method. We consider a black-box poisoning attack against the PPO-based DRL agent. To this end, we use a weak adversary attack as in \cite{wu2021reinforcement} to attack the system.
Suppose the attacker determines to attack the time step $t$, it generates an arbitrary state $\hat{s}_t$ and the associated reward function $\hat{r}(\hat{s}_t,.)$. When the agent observes the altered state $\hat{s}_t$, it applies action $a_t$ and observes $\hat{r}(\hat{s}_t,a_t)$, rather than $r(s_t,a_t)$.

\begin{figure}
\vspace{-1.5em}
\centering
\subfigure[]{
\includegraphics[scale=0.36, trim={0cm 0 0 0}]{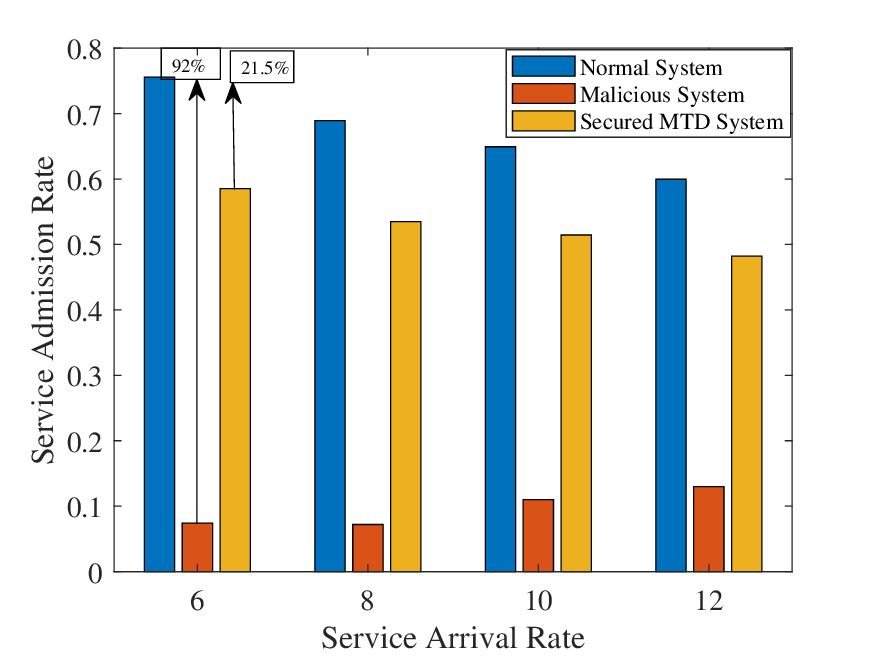}\label{fig:c14a}
}\hspace{-0.25cm}
\subfigure[]{
\includegraphics[scale=0.36, trim={0 0 0 0}]{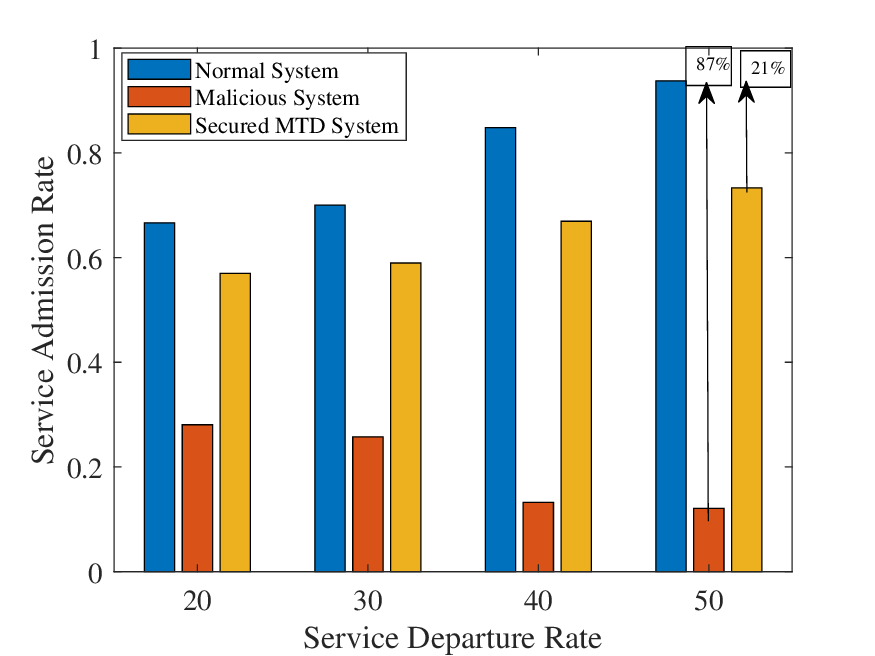}\label{fig:c14b}
}
\caption{Service admission rate vs. (a) mean service arrival rate and (b) mean service departure rate.}
\label{fig:c14}
\end{figure}
Therefore, we assume that in each time step, the state of the system, which is the remaining memory and the service arrival rate of two services, is perturbed. In our simulations, we altered the service arrival rates of two services and converted them to the uniform random variable between zero and the service arrival rate. Therefore, we blocked part of service arrival rates in these simulations based on the weak adversary attack in \cite{wu2021reinforcement}.

\subsubsection{MTD technique}
\label{sec:mtd-technique}
To tackle the adversarial attack issue, we adopt the MTD approach, where the defender has multiple configurations for the ML models. In this scenario, as shown in Fig \ref{fig:c15}, we use four different PPO models with varying configurations for learning. We assume that the adversarial attacker can randomly affects one of these models during the training.
After the models are trained, a random model is selected among the four models to run each input and returns the output generated by that model.
Thanks to the dynamicity introduced by the proposed MTD method, attackers will have less impact on the system because they attack one of the models and do not know which model is selected.

In this scenario, we delve into the O-RAN near-RT RIC architecture, specifically employing the AI/ML approach, notably the PPO model, for resource allocation. The RIC layer, constituting the new AI/ML controller within the O-RAN system, plays a pivotal role in the service admission control and resource allocation. As elucidated in the O-RAN white papers, RL methods find implementation within the near-RT RIC for the resource allocation. In this context, we explore the integration of MTD for fortifying the system. To accomplish this, we trained four distinct models, each configured as an individual xApp in the near RT RIC.

\subsubsection{Performance Results}
\label{sec:performance}

To evaluate the efficiency of the PPO-based dynamic service admission control solution and the effectiveness of the proposed MTD method in withstanding adversarial attacks against DRL, we consider three scenarios. The first scenario is a regular system without any attack with one PPO model. The second scenario is where we have an attack in the system with one PPO model.
In the third scenario, we use the proposed MTD technique with four PPO models. For the three scenarios, the average service admission rate is measured in terms of the mean service arrival rate and the mean service departure rate. Fig.~\ref{fig:c14a} and Fig.~\ref{fig:c14b} report the comparative results. It is observed that the service admission rate of the system decreases with the increase of the service arrival rate, which is attributed to the limited available resources. Furthermore, as the service departure rate increased, the service admission rate increased due to the release of memory. We can also notice a significant enhancement in the system's performance under adversarial attacks after using the MTD technique. Fig.~\ref{fig:c14a} shows that the secured MTD system experienced only $21.5\%$ lower admission rate under adversarial attack, compared to $91\%$ drop-in admission rate when the system is not secured. Similar observations hold true in Fig.~\ref{fig:c14b}, where we can see that the secured MTD system limited the attacker's impact to $21\%$ decrease in the admission rate, compared to $87\%$ without protection from adversarial attacks.
\vspace{-1em}
\subsection{LLM-based XAI Robust AI/ML in O-RAN} 
\label{llm}
\textcolor{black}{In a previous scenario, the AI/ML component responsible for service admission control was managed using the PPO model. We assumed a weak adversarial attack was in play. To diagnose and explain this unusual behavior, an LLM XAI system could take action. For example, the LLM could analyze the model's decision-making process and generate a plain-language report: "The service admission model has rejected 15 devices in the last 15 minutes, a significant difference from its normal pattern of one rejection per 15 minutes."}

\textcolor{black}{The LLM system employs XAI techniques to identify the malicious model. Using the Isolation Forest technique, an unsupervised ML algorithm for anomaly detection, the system can detect outlier data based on features such as mean and variance. The LLM then explains these anomalies in a human-readable format. This insight enables the O-RAN system to quickly recognize malicious interference with the PPO model. An immediate investigation is recommended to confirm the nature of the detected anomaly and take steps to remove that model from the system.}

\textcolor{black}{By leveraging the capabilities of the LLM-based XAI system, network operators can gain a deeper understanding of the underlying issues affecting AI/ML-driven service admission control. This will ensure that the integrity and security of the O-RAN system are maintained.}
\subsubsection{System Scenario}
In this scenario, we demonstrate how the LLM system can analyze data and translate it into human-readable language to assist in detecting and mitigating attacks within the MTD system. This represents an advanced MTD system that integrates the LLM model and XAI to analyze and clarify attacks, subsequently removing the affected model from the MTD system. Suppose one of the four models is targeted in an attack. When the system selects this xApp, the data pattern for service admission differs from that of other xApps (i.e., service admission is notably lower for this specific xApp compared to others). The LLM system can analyze the data pattern, identify the attacked model based on the pattern, describe it in human-readable language, and then request action, which could be performed by either the system operator or the SMO, to remove the specific xApp from the O-RAN system \cite{dave2024integrating}.
\subsubsection{Proposed method}
\textcolor{black}{We studied Fig. \ref{fig:c14} (where service arrival rate is 12) whenever one of the 4 trained models was attacked. We used GPT-4's data analyst with isolation forest to spot unusual patterns in the outputs of these four models over time.}

\textcolor{black}{We provided the data to GPT-4 for malicious activity detection within the system. The service admission rates for models x1, x2, and x4 were similar, averaging around 60\%, whereas model x3 averaged approximately 15\%.}

\textcolor{black}{The results reveal significant differences and potential issues among the series analyzed. Series x1 and x4 display consistent values with moderate variation typical of time series data. Series x2 shows higher peaks (e.g., 63) and slightly more variability, which seems contextually normal. In contrast, series x3 stands out with consistently lower and less varied values. Identified as an anomaly by the Isolation Forest algorithm, x3 exhibits significantly lower mean and variance compared to x1, x2, and x4. This deviation suggests poisoning attack or any error in the system. Further investigation, including system log reviews, configuration checks, or security audits, is essential to identify and address potential malicious activity or technical faults in x3.}

\section{Conclusion} 
\label{sec:conclusion}
\noindent
This paper investigated the threat landscape applying to the emerging O-RAN architecture. After briefly introducing the O-RAN architecture, we discussed the main vulnerabilities and threats against the O-RAN system. As a result, and in view of bolstering the security posture of O-RAN, we recommended and discussed the potential of three emerging approaches, namely the ZT concept, blockchain technology, LLM and MTD paradigm. A proof of concept has been presented, showing the effectiveness of MTD in strengthening the robustness of DRL models to adversarial poisoning attacks.
Moreover, we also studied the effect of LLM in detecting the attack in the O-RAN AI/ML system using an example.
Despite the merits of the four advocated approaches, their adoption in securing O-RAN is still facing different challenges, including (i) enabling continuous monitoring and assessment of risks; a key requirement for ZT, with reduced impact on network performances; (ii) solving the scalability, performance and privacy challenges for blockchain; (ii) developing advanced MTD strategies that can meet the desired trade-off between robustness, performance and moving cost; and (iii) introducing the LLM system to automate the system, bringing XAI techniques, decreasing the risk of threats in AI/ML techniques.


\end{document}